\documentclass[12pt]{article}
\def\slashchar#1{\setbox0=\hbox{$#1$}           
   \dimen0=\wd0                                 
   \setbox1=\hbox{/} \dimen1=\wd1               
   \ifdim\dimen0>\dimen1                        
      \rlap{\hbox to \dimen0{\hfil/\hfil}}      
      #1                                        
   \else                                        
      \rlap{\hbox to \dimen1{\hfil$#1$\hfil}}   
      /                                         
   \fi}                                         %
\newcommand{\blank}{}
\renewcommand{\theequation}{\blank \arabic{equation}}
\newcommand{\startappendix}{%
    \appendix
    \setcounter{equation}{0}
    \renewcommand{\blank}{\Alph{section}.}%
  }
\newcounter{dummy}{}
\newcommand{\letters}{%
    \setcounter{dummy}{\value{equation}}
    \renewcommand{\thedummy}{\blank \arabic{dummy}}
    \renewcommand{\theequation}{\thedummy\alph{equation}}
    \refstepcounter{dummy}
    \setcounter{equation}{0}%
 }
\newcommand{\noletters}{%
    \setcounter{equation}{\value{dummy}}
    \renewcommand{\theequation}{\blank\arabic{equation}}%
  }
\newenvironment{mathletters}{\letters}{\noletters}
\newcommand{\half}{\frac 12}
\newcommand{\sgn}{\mathop{\mathrm{sgn}}}
\newcommand{\diag}{\mathop{\mathrm{diag}}}
\begin{document}
\title{Dirac operator on the Riemann sphere}
\author{A.~A.~Abrikosov,~jr.\thanks{E-mail:
{\sc persik@vitep1.itep.ru}} \\[2mm]
{\em ITEP, B.~Cheremushkinskaya~25,} \\
{\em 117 259, Moscow, Russia}}
\date{}
\maketitle
\begin{abstract}
We solve for spectrum, obtain explicitly and study group
properties of eigenfunctions of Dirac operator on the Riemann
sphere $S^2$. The eigenvalues $\lambda$ are nonzero integers. The
eigenfunctions are two-component spinors that belong to
representations of $SU(2)$-group with half-integer angular momenta
$l = |\lambda| - \half$. They form on the sphere a complete
orthonormal functional set alternative to conventional spherical
spinors. The difference and relationship between the spherical
spinors in question and the standard ones are explained.
\end{abstract}

\smallskip

\noindent {\sc keywords:} \qquad Dirac operator, spherical
functions, spherical spinors, orthogonal polynomials, monopole
harmonics.
\smallskip

\noindent arXiv:~hep-th/0212134

\section*{Introduction}

The question of spectrum and eigenstates of Dirac operator on the
Riemann sphere $S^2$ is a well posed mathematical problem that may
be solved exactly. The resulting family of orthogonal
two-component spinors plays the role of orthogonal spherical
functions in problems with half-integer momenta. Therefore the
solutions are of fundamental value.

The eigenspinors were first found by Newman and Penrose
\cite{Newman/Penrose} and the relation with SU(2) matrix elements
was deduced later in
\cite{Goldberg/Macfarlane/Newman/Rohrlich/Sudarshan}. These
eigenspinors also reappear from time to time under the name of
``monopole harmonics'', see, for instance \cite{Dray}. A general
construction of Dirac operator eigenfunctions on $N$-dimensional
spheres was given in paper \cite{camporesi/higuchi} and the
two-dimensional case was specially addressed in
\cite{varilly/book} (see also \cite{varilly/lectures}).

One must mention several physical situations where the solutions
may find application. The first involves the so-called spectral
boundary conditions. Those have been introduced by Atiah, Patodi
and Singer \cite{APS} and are widely used now in studies of
spectral properties of operators in the limited space
\cite{Eguchi/Gilkey/Hanson}. Being compatible with chiral symmetry
these boundary conditions are perfectly suited for investigations
of chirality breaking in quantum field theory.

The spectral boundary conditions are defined in terms of
eigenvalues of Dirac operator on the boundary. Thus our problem
bears a direct relation to the (spherical) spectral bag that
allows to hold quarks while preserving the chiral symmetry.

Another application refers to physics of electrons in substances
called fullerenes. Their molecules consist of carbon atoms located
in vertices of polyhedra or on the surface of clusters with a
spheroidal crystallographic structure. The most abundant and the
most spherical is the $\mathrm{C}_{60}$ molecule (known also as
Buckminsterfullerene or buckyball) that looks like a standard
soccer ball. The interest to electronic structure of fullerenes
was heated by their unconventional superconducting properties with
transition temperatures higher than those of the classic
superconductors (reaching up to 60~K!). In the continuum limit the
conducting electrons of fullerene molecules obey the Dirac
equation \cite{Gonzalez/Guinea/Vozmediano} and our results are
pertinent for the continuous description of electron states in
fullerenes.

Our present goal was to implement an explicit classical approach
to eigenstates of the Dirac operator with the intention to make
them a tool that could be later employed in physical problems. A
short presentation of the topic was given in \cite{Leipzig2001}.
In principle one could extract the required information from the
already cited papers but this would take some effort.

It was shown in  \cite{camporesi/higuchi} that for Dirac operator
on the $N$-dimensional sphere one may separate the dependence on
principal polar angle and solve the obtained second order
differential equation. The result was written in terms of Jacobi
polynomials. This led to a recurrent formula that related spinor
spherical functions on $S^N$ to those on $S^{N-1}$. The subsequent
reduction (quite conceivable for $S^2$ though) was left to the
reader.

The two-dimensional case was considered  as an example in
\cite{varilly/book} (Chapter~9) and may be also found in
\cite{varilly/lectures}. However there the problem was elegantly
solved with the help of complex coordinates on the Riemann sphere
whereas the transformation to standard spherical coordinates was
left aside.

With practical applications in mind we shall try to make the paper
self-contained. It has the following structure.
Section~\ref{general} introduces the notation and expounds on
general properties of Dirac operator on the Riemann sphere. The
spectrum and eigenfunctions themselves are investigated in
Section~\ref{Dirac}. We start from the eigenvalues
(Sect.~\ref{ev-problem}), analyse $SU(2)$ properties of solutions
(Sect.~\ref{SU(2)}), then write them out explicitly
(Sect.~\ref{Upsilon}) and check their properties with respect to
time-reversal (Sect.~\ref{c-conj}). Section~\ref{Ups-vs-Om}
explains the link between the newly found spherical spinors and
the conventional ones \cite{Akhiezer/Berestetsky}. After reminding
the spinor transformation rules (Sect.~\ref{spinor-transform}) we
convert our eigenspinors to Cartesian coordinates and derive the
relationship in question (Sect~\ref{Ups/Om}). The conclusion is
followed by two technical Appendices dwelling on properties of
Jacobi polynomials (Appendix~\ref{Jacobi}) and spherical functions
(Appendix~\ref{Sph-func}).

\section{Generalities \label{general}}

We shall start with introducing spherical coordinates and writing down the
Dirac operator for free massless fermions on the Riemann sphere $S^2$. Then
we shall show that Dirac operator has has no zero eigenvalues on the sphere.

The sphere of unit radius $S^2$ may be parameterized by two
spherical angles $q^1=\theta$, $q^2=\phi$ that are related to
Cartesian coordinates $x,\, y,\, z$ as follows:
\begin{equation} \label{sph-coord}
 x = \sin \theta \cos \phi ;   \quad
 y = \sin \theta \sin \phi ;   \quad
 z = \cos \theta .
\end{equation}
The metric tensor and the natural diagonal zweibein on the sphere are well
known:
\begin{equation}\label{zweibein}
 g_{\alpha \beta }=\diag (1,\, \sin^2\theta);
    \qquad
 e_\alpha^a= \diag (1,\, \sin \theta);
    \quad
 g_{\alpha \beta }
    = \delta_{ab}\, e_\alpha ^a\, e_\beta^b.
\end{equation}
From here on we shall denote zweibein and coordinate indices by
Latin and Greek letters respectively. Multiplication by the
zweibein converts the two types of indices into each other:
\begin{equation}\label{a->alpha}
  A_a = e^\alpha_a \, A_\alpha;
    \qquad
  A^a = e_\alpha^a \, A^\alpha;
    \qquad
 e_\alpha^a \, e^\alpha_b = \delta^a_b.
\end{equation}

In curvilinear coordinates the ordinary Euclidean partial
derivatives must be replaced by covariant ones. For quantities
with Latin indices those
include the spin connection $R_{\alpha \,b}^a$%
\footnote{Since there is actually no difference between upper and lower
Latin indices the second of eqs.~(\ref{covDA}) is nothing but the skew
symmetry of $\hat{R}$.}:
\begin{equation}\label{covDA}
  D_\alpha A^a =
    \partial_\alpha A^a + R_{\alpha \,b}^a\, A^b
    \qquad \mathrm{and} \qquad
  D_\alpha A_a =
    \partial_\alpha A_a - A_b \, R_{\alpha \,a}^b .
\end{equation}
In our case the nonzero components of spin connection are:
\begin{equation} \label{R-S2}
R^{12}_\phi = - R^{21}_\phi = - \cos \theta.
\end{equation}

Spinors in two dimensions have two components and the role of Dirac matrices
belongs to Pauli matrices: $\gamma^a \rightarrow (\sigma_x,\, \sigma_y)$
where
\begin{equation}\label{Pauli}
  \sigma_x = \left(
\begin{array}{cc}
  0 & 1 \\
  1 & 0
\end{array} \right); \qquad
\sigma_y = \left(
\begin{array}{rr}
  0 & -i \\
  i & 0
\end{array} \right); \qquad
\sigma_z = \left(
\begin{array}{rr}
  1 & 0 \\
  0 & -1
\end{array} \right).
\end{equation}

Covariant derivatives of 2-spinors are also expressed in terms of the spin
connection:
\begin{equation}\label{Dpsi}
  \nabla_\alpha \psi =
    \partial_\alpha \psi + \frac i4 R_\alpha^{ab}\, \sigma_{ab}\, \psi.
\end{equation}
Here $\sigma_{ab}$ are the rotation generators for spin-$\frac 12 $ fields:
\begin{equation} \label{sigma12}
  \sigma _{1\,2} = - \sigma _{2\,1} =
    - \frac i2 [\gamma_1,\, \gamma_2] = \sigma_z.
\end{equation}
A standard calculation gives for covariant derivatives:
\begin{equation} \label{nablaS2}
 \nabla_\theta  =  \partial_\theta
    \qquad \mathrm{and}  \qquad
  \nabla_\phi  = \partial_\phi - \frac {i\sigma_z}2 \cos \theta.
\end{equation}

Now we may define the Dirac operator. In spherical coordinates it is given
by the convolution of covariant derivatives in spinor representation with
the zweibein and $\sigma$-matrices:
\begin{equation}\label{Dirac-S2}
  -i \hat{\nabla} = -i\, e^{\alpha\,a} \sigma_a \nabla_\alpha.
\end{equation}
With the help of definitions (\ref{zweibein}, \ref{Pauli}, \ref{nablaS2}) it
is straightforward to obtain:
\begin{equation}\label{Dirac/sph}
  -i\,\hat{\nabla} =
    -i \sigma_x
    \left(\partial_\theta + \frac{\cot\theta}2 \right)
    -\frac{i \sigma_y}{\sin\theta}\, \partial_\phi .
\end{equation}

Finally let us show that Dirac operator on the sphere has no zero
eigenvalues. This is a general property of manifolds with positive
curvature known as Lichnerowicz theorem
\cite{Berline/Getzler/Vergne}. Consider the square of Dirac
operator $(-i \hat{\nabla})^2$. Obviously if $-i \hat{\nabla}$ had
a zero eigenvalue then $(-i \hat{\nabla})^2$ must have it either.
Let us split the product of $\sigma$-matrices in
$(\hat{\nabla})^2$ into symmetric and antisymmetric parts
 $\sigma^\alpha \sigma^\beta = \half \{\sigma^\alpha,\, \sigma^\beta\} +
    \half [\sigma^\alpha,\, \sigma^\beta]$.
Implementing the relation between commutators of covariant derivatives and
curvature we obtain:
\begin{equation}\label{D-L=comm}
    (-i \hat{\nabla})^2 + \nabla_F^2 =
    -\frac i2 \sigma^{\alpha \beta}\, [\nabla_\alpha , \, \nabla_\beta] =
    \frac 14 \, R^{\alpha\beta}_{\alpha\beta}.
\end{equation}
Here $R^{\alpha\beta}_{\alpha\beta}$ is the trace of Riemann curvature
tensor and $\nabla_F^2$ is the covariant Laplace operator in fundamental
representation of the $SU(2)$-group:
\begin{equation}\label{Laplace}
  \nabla_F^2 = g^{\alpha\beta}\,
    (\nabla_\alpha \nabla_\beta - \Gamma^\gamma_{\alpha\beta}\nabla_\gamma)
    = \frac 1{\sqrt{g}}\,
    \nabla_\alpha\, g^{\alpha\beta}\sqrt{g}\, \nabla_\beta ,
\end{equation}
where $\nabla_{\alpha,\, \beta,\, \gamma}$ are covariant derivatives in
fundamental representation (\ref{Dpsi}), $\Gamma^\gamma_{\alpha\beta}$ is
the Christoffel symbol and $g$ is the determinant of metric,
 $g = \det \|g_{\alpha\beta} \| = \sin^2 \theta$.

The result (\ref{D-L=comm}) is called the Lichnerowicz formula. In the case
of sphere $R^{\alpha\beta}_{\alpha\beta} = 2$ so that
\begin{equation}\label{Dsqr-Lapl}
  (-i \hat{\nabla})^2 + \nabla^2 = \half.
\end{equation}
(certainly  this may be proved directly). The positive difference is brought
about by the curvature of Riemann sphere. The covariant Laplacian is a
non-positive operator $\nabla^2 \leq 0$. Hence, by virtue of
(\ref{Dsqr-Lapl}), $(-i \hat{\nabla})^2$ must be strictly positive and Dirac
operator $-i \hat{\nabla}$ has no zero eigenvalues.

In conclusion I would like to present the explicit formulae for the square
of Dirac operator and the fundamental Laplace operator on the unit sphere:
\begin{mathletters}\label{Dsqr}
\begin{eqnarray}
  (-i \hat{\nabla})^2 & = &
    -\frac 1{\sin \theta} \partial_\theta\, \sin \theta\, \partial_\theta -
    \frac{\partial_\phi^2}{\sin^2\theta} +
    i \sigma_z \frac{\cos \theta}{\sin^2\theta}\partial_\phi +
    \frac 14 + \frac 1{\sin^2 \theta};
\label{Dsqra} \\
  - \nabla_F^2 & = &
    -\frac 1{\sin \theta} \partial_\theta\, \sin \theta\, \partial_\theta -
    \frac{\partial_\phi^2}{\sin^2\theta} +
    i \sigma_z \frac{\cos \theta}{\sin^2\theta}\partial_\phi -
    \frac 14 + \frac 1{\sin^2 \theta}.
\label{Dsqrb}
\end{eqnarray}
\end{mathletters}
Obviously the difference between the two agrees with (\ref{Dsqr-Lapl}).

\section{Dirac operator on the sphere\label{Dirac}}

Now we shall study the eigenvalue problem for Dirac operator on
the sphere $S^2$. First we shall derive the differential equation
for spinor components, find eigenvalues and the general form of
solutions. Then we shall study properties of the eigenfunctions
that arise from the $SU(2)$-invariance of the problem and classify
the solutions. Finally we shall present them explicitly and
discuss their behaviour under the complex conjugation and
time-reversal.

\subsection{The eigenvalue problem\label{ev-problem}}

Eigenfunctions of the Dirac operator (\ref{Dirac/sph}) are two-component
spinors that satisfy the eigenvalue equation:
\begin{equation}\label{DiracS2}
 -i\,\hat{\nabla}
    \left(\begin{array}{c}
      \alpha_\lambda (\theta,\, \phi)  \\
      \beta_\lambda (\theta,\, \phi)
    \end{array}\right) =
    \lambda\, \left(\begin{array}{c}
      \alpha_\lambda (\theta,\, \phi) \\
      \beta_\lambda (\theta,\, \phi)
    \end{array}\right).
\end{equation}

This system of first order partial differential equations in
$\alpha$ and $\beta$ allows separation of variables. The first
thing is to isolate the $\phi$-dependence by expanding the spinors
into Fourier series:
\begin{equation}\label{e(imfi)}
  \left(\begin{array}{c}
      \alpha_\lambda (\theta,\, \phi) \\
      \beta_\lambda (\theta,\, \phi)
    \end{array}\right) = \sum_m
    \frac{\exp i\, m\, \phi}{\sqrt{2\pi}}
    \left(\begin{array}{c}
      \alpha_{\lambda m} (\theta) \\
      \beta_{\lambda m} (\theta)
    \end{array}\right);
    \qquad
    m = \pm \half,\, \pm \frac 32 \dots
\end{equation}
where $m$ are half-integers since we work with the spin-$\half$ field. This
converts Eqn.~(\ref{DiracS2}) into
\begin{mathletters}\label{Phi-harm}
\begin{eqnarray}
  -i \left(\partial_\theta + \frac{\cot\theta}2
    + \frac m{\sin\theta}\right)\,
    \beta_{\lambda m}(\theta) & = &
    \lambda\, \alpha_{\lambda m}(\theta) ;
 \label{Phi-harma} \\
  -i \left(\partial_\theta + \frac{\cot\theta}2
    - \frac m{\sin\theta}\right)\,
    \alpha_{\lambda m}(\theta) & = &
    \lambda\, \beta_{\lambda m}(\theta) .
 \label{Phi-harmb}
\end{eqnarray}
\end{mathletters}
By analogy with the ordinary quantum mechanics the number $m$ may be called
the projection of angular momentum onto the polar axis.

Separate equations for spinor components $\alpha$ and $\beta$ may be
obtained by taking the square of Dirac operator (\ref{Dsqra}):
\begin{equation}\label{dAlembS2}
  \left[- \frac 1{\sin \theta}\, \partial_\theta\, \sin \theta\, \partial_\theta
    +\frac{m^2}{\sin^2 \theta} - \sigma_z \frac{m\, \cos \theta}{\sin^2 \theta}
    + \frac 14 + \frac 1{4\sin^2 \theta}\right]\,
    \left(\begin{array}{c}
       \alpha_{\lambda m}  \\
       \beta_{\lambda m}
    \end{array}\right) =
    \lambda^2 \, \left(\begin{array}{c}
       \alpha_{\lambda m}  \\
       \beta_{\lambda m}
    \end{array}\right).
\end{equation}
Mark that because of the $\sigma_z$ present in the third term the equations
for upper and lower components are different. As may be seen from
eqn.~(\ref{Dsqr-Lapl}) the difference arises already at the level of Laplace
operator in fundamental representation (\ref{Dsqrb}).

The further simplification comes from the change of variables
 $x =\cos \theta$, $x \in [-1,\, 1]$, that converts (\ref{dAlembS2})
into the generalized hypergeometric equation:
\begin{equation}\label{ODEf}
  \left[\frac d{dx}\, (1-x^2)\, \frac d{dx} -
    \frac{m^2 - \sigma_z\, m\, x +\frac 14}{1-x^2}\right] \left(
    \begin{array}{c}
      \alpha_{\lambda m}(x) \\
      \beta_{\lambda m}(x)
    \end{array}\right) =
    - \left(\lambda^2 -\frac 14\right)\, \left(
    \begin{array}{c}
      \alpha_{\lambda m}(x) \\
      \beta_{\lambda m}(x)
    \end{array}\right).
\end{equation}
Take the notice that the replacement $x \rightarrow -x$ (or $m
\rightarrow -m$) is equivalent to trading $\alpha$ for $\beta$.
Thus the upper and lower spinor components are conjugate with
respect to mirror reflection.

Equation (\ref{ODEf}) is singular at the poles of the sphere $x = \pm 1$.
After the redefinition of the unknowns
\begin{equation}\label{mvalued}
  \left(
    \begin{array}{c}
      \alpha_{\lambda m}(x) \\
      \beta_{\lambda m}(x)
    \end{array}\right) = \left(
    \begin{array}{c}
      (1-x)^{\half\left|m - \half\right|} (1+x)^{\half\left|m + \half\right|}
    \, \xi_{\lambda m}(x) \\
      (1-x)^{\half\left|m + \half\right|} (1+x)^{\half\left|m - \half\right|}
    \, \eta_{\lambda m}(x)
    \end{array}\right),
\end{equation}
we arrive to the separate equations of hypergeometric type in
 $\xi_{\lambda m}$ and $\eta_{\lambda m}$:
\begin{equation}\label{J-eqns}
  \left\{(1-x^2)\frac{d^2}{dx^2} +
    \left[\frac m{|m|}\sigma_z - (2|m| +2)\, x\right] \frac d{dx}
    -m\,(m+1) + \left(\lambda^2 - \frac 14\right) \right\} \left(
    \begin{array}{c}
      \xi_{\lambda m} \\
      \eta_{\lambda m}
    \end{array} \right) = 0.
\end{equation}

In order that these equations had on the interval $x \in [-1,\, 1]$ square
integrable solutions $\lambda$ must fulfill the condition (see
(\ref{J-eq/def})--(\ref{lambda}))
\begin{equation}\label{lmb=m+n+}
  \lambda^2 = \left(n + |m| + \half\right)^2,
\end{equation}
with non-negative integer $n \geq 0$. Thus $\lambda$ are nonzero integers
\begin{equation}\label{intSpct}
  \lambda = \pm 1,\, \pm 2 \dots,
\end{equation}
and indeed the Dirac operator has no zero-modes. Equations
(\ref{J-eqns}) are Jacobi-type equations and their solutions are
Jacobi polynomials of the $n$-th order \cite{Nikiforov/Uvarov,
Bateman/Erdelyi}:
\begin{equation}\label{J-sln's}
  \left(
    \begin{array}{c}
      \xi_{\lambda m}(x) \\
      \eta_{\lambda m}(x)
    \end{array} \right) =
  \left(
    \begin{array}{c}
      C^{mn}_\alpha\, P^{(|m-\half|,\, |m+\half|)}_{n}(x) \\
      C^{mn}_\beta\,  P^{(|m+\half|,\, |m-\half|)}_{n}(x)
    \end{array} \right) .
\end{equation}
A brief review of Jacobi polynomials may be found in Appendix~A. Properties
of solutions (\ref{J-sln's}) will be discussed in Section~\ref{Upsilon}.

The constants $C^{mn}_\alpha,\, C^{mn}_\beta$ may be fixed by substituting
the solutions into equation (\ref{DiracS2}) and normalizing the obtained
functions. Let us first find the relation between $C_\alpha$ and $C_\beta$.
The Dirac operator is sensitive to the sign of $m$. Applying it to functions
(\ref{mvalued}) at $m>0$ gives for the Fourier components (the same can be
done for $m<0$):
\begin{equation}\label{C-eqns}
  -i \hat{\nabla}\left(
    \begin{array}{c}
    \alpha_{\lambda m}\\
    \beta_{\lambda m}
  \end{array}\right) = i (1-x^2)^{\frac m2}\, \left(
    \begin{array}{r}
      - \left(\frac{1+x}{1-x}\right)^\frac 14 \, \left[
        \left(m + \half \right) \eta_{\lambda m}
        - (1-x) \frac d{dx} \eta_{\lambda m} \right] \\
      \left(\frac{1-x}{1+x}\right)^\frac 14 \, \left[
        \left(m + \half \right) \xi_{\lambda m}
        + (1+x) \frac d{dx} \xi_{\lambda m} \right]
    \end{array}\right)
\end{equation}
(Note that coefficients $C_{\alpha,\, \beta}$ were absorbed in functions
$\xi$, $\eta$.) Substituting this into the eigenvalue equation
(\ref{DiracS2}) written in terms of $\xi$ and $\eta$ and using identities
(\ref{ab-ids}) of the Appendix we find the condition on $C^{mn}_\alpha,\,
C^{mn}_\beta$ which, after generalization to negative $m$, takes the form:
\begin{equation}\label{C-solved}
  i \left(n + |m| +\half\right)\left(
\begin{array}{r}
  C^{mn}_\alpha \\
  - C^{mn}_\beta
\end{array}\right) = \lambda \sgn m \, \left(
\begin{array}{c}
  C^{mn}_\beta \\
  C^{mn}_\alpha
\end{array}\right).
\end{equation}
Provided that $\lambda$ is given by (\ref{lmb=m+n+}) this equation has
nonzero solutions and the ratio of constants  $C_\alpha / C_\beta$ is
defined by the sign of the product $m\lambda$:
\begin{equation}\label{C/C}
  C^{mn}_\beta = i\, C^{mn}_\alpha \sgn (m \, \lambda).
\end{equation}

The absolute values of the constants can be found from the normalization
conditions,
\begin{equation}\label{e-norm}
  \int_0^{2\pi} d\phi \int_0^{\pi}
    \left[|\alpha_\lambda(\theta,\, \phi)|^2 +
    |\beta_\lambda(\theta,\, \phi)|^2\right]\,
     \sin \theta \, d\theta =1.
\end{equation}
This immediately refers us to the norms of Jacobi polynomials with the
result:
\begin{equation}\label{C-mod}
  | C_\alpha^{mn} | = | C_\beta^{mn} | =
    \frac{\sqrt{n!\, (n+2m)!}}{2^{m+\half}\, \Gamma(n+m+\half)}.
\end{equation}

The last condition determines the eigenfunctions up to a complex phase that
will be fixed in Section~\ref{c-conj}. Presently we shall not write the
solutions explicitly leaving this till Section~\ref{Upsilon}.

The last remark due here concerns an apparent contradiction
between the integer spectrum (\ref{intSpct}) and formula
(\ref{Dsqr-Lapl}). The latter implies that spectra of the two
operators (\ref{Dsqr}) are similar but to the shift $\half$. Hence
according to formula (\ref{intSpct}) the lowest eigenvalue of the
(minus) covariant Laplace operator $-\nabla^2_F$ is $\half$ and
not zero as one could naively expect by analogy with scalar
fields. The confusion is readily resolved if one recalls that the
aforementioned integer spectrum was obtained for spinors. Unlike
scalars those satisfy antiperiodic boundary conditions $\alpha,\,
\beta (0) = -\alpha,\, \beta (2\pi)$ and are expanded in
half-integer Fourier harmonics. Therefore although formally
 $- \nabla^2$ and $-\hat{\nabla}^2$
differ only by the numerical constant in fact both spectra and
eigenfunctions for scalars and spinors are different.

\subsection{The $SU(2)$ algebra} \label{SU(2)}

Let us show that the Dirac operator on the sphere $S^2$ is invariant under
transformations of the $SU(2)$ group. The Weyl~---Cartan set of generators
looks as follows:
\begin{mathletters}\label{L-ops}
\begin{eqnarray}
  \hat{L}_z & = &
    - i \frac \partial{\partial \phi}; \label{L-ops/a}\\
  \hat{L}_+ & = &
    \hphantom{-} e^{\hphantom{-} i \phi}
    \left(\frac \partial{\partial \theta}
    + i\, \cot \theta \frac \partial{\partial \phi}
    + \frac{\sigma_z}{2\, \sin \theta} \right); \label{L-ops/b}\\
  \hat{L}_- & = &
   - e^{- i \phi}
    \left(\frac \partial{\partial \theta}
    - i\, \cot \theta \frac \partial{\partial \phi}
    - \frac{\sigma_z}{2\, \sin \theta} \right).  \label{L-ops/c}
\end{eqnarray}
\end{mathletters}
These operators satisfy the standard commutation relations of the $SU(2)$
algebra:
\begin{mathletters}\label{L-comm}
\begin{eqnarray}
  \left[ \hat{L}_z ,\, \hat{L}_+ \right] & = &
    \hphantom{-} \hat{L}_+ ;  \label{L-comm/a}\\
  \left[ \hat{L}_z ,\, \hat{L}_- \right] & = &
    - \hat{L}_- ;  \label{L-comm/b}\\
  \left[ \hat{L}_+ ,\, \hat{L}_- \right] & = &
    \, 2 \hat{L}_z .  \label{L-comm/c}
\end{eqnarray}
\end{mathletters}

In our case representations of $SU(2)$-group are characterized by
half-integer highest weight $l$ (see (\ref{l=m+n}). It will be
shown in Section~\ref{Ups/Om} that it corresponds to the total
angular momentum of the fermion state. The Casimir operator
$\hat{L}^2 $ takes the value
\begin{equation}\label{Lsqr}
  \langle l \mid \hat{L}^2 \mid l\rangle   =
    \langle l \mid \hat{L}_z^2 +
    \half(\hat{L}_+ \hat{L}_- + \hat{L}_- \hat{L}_+)\mid l\rangle
    = l(l+1).
\end{equation}
The basis vectors of the representation are classified by
half-integer projection of angular momentum onto the polar axis
that varies in the range $m = l_z = -l,\dots\, l$. The number of
basis vector in the representation is $2l+1$. The raising and
lowering operators $\hat{L}_+$ and $\hat{L}_-$ change the
projection by unity
 $\hat{L}_\pm \mid l,\, m \rangle \propto \mid l,\, m \pm 1 \rangle$.

A direct check proves that generators (\ref{L-ops}) commute with
spherical Dirac operator (\ref{Dirac/sph}):
\begin{equation}\label{nabla-comm}
  \left[-i\hat{\nabla},\, \hat{L}_z \right] =
    \left[-i\hat{\nabla},\, \hat{L}_+ \right] =
    \left[-i\hat{\nabla},\, \hat{L}_- \right] = 0.
\end{equation}
This means that the Dirac operator is $SU(2)$ invariant and its
eigenfunctions may be classified according to their $SU(2)$ transformation
properties. Moreover, the action of raising and lowering generators
$\hat{L}_+$ and $\hat{L}_-$ transforms its eigenfunctions into the
eigenfunctions with the same $\lambda$ but greater or smaller $m$
respectively.

The square of Dirac operator (\ref{Dsqra}) and Casimir operator (\ref{Lsqr})
may be diagonalized simultaneously
\begin{equation}\label{nabla/L}
  - \hat{\nabla}^2 = \hat{L}^2 + \frac 14,
\end{equation}
and their eigenvalues are interrelated:
\begin{equation}\label{L-sqr:mn}
  \langle \lambda,\, n | \hat{L}^2 | \lambda,\, n \rangle =
    l(l+1) =
     \lambda^2 - \frac 14 = (n + |m|)(n + |m| + 1).
\end{equation}
We conclude that the proper values of angular momentum are indeed
half-integers related to the eigenvalues of Dirac operator
\begin{equation}\label{l=m+n}
  l = |\lambda| - \half = n + |m| = \half,\, \frac 32,\, \frac 52 \dots
\end{equation}

The final remark concerns reducibility of our representation. It
is well known that spinor representations in even dimensions are
reducible and one may separate the left and right spinor
components. One may see that formally this is the case. Operators
(\ref{L-ops}) are diagonal, so the upper and lower spinor
components behave with respect to the $SU(2)$-group like two
singlets with equal momenta $l$. However diagonalizing the Dirac
operator requires joining them into the pair of doublets with
$\lambda = \pm (l+\half)$.

\subsection{The spinor spherical functions} \label{Upsilon}

Here we shall show how the formerly obtained functions
(\ref{J-sln's}) can be grouped into $SU(2)$ multiplets. The
multiplets can be constructed in the standard manner,
\emph{i.~e\/} by successive application of the lowering operator
$\hat{L}_-$ to the highest weight vector, namely to the function
with the maximum projection $m = l$ of angular momentum. Let us
pass right to the result.

It is quite natural to use instead of $n$ the angular momentum
 $l = n + |m|$. Let us introduce the integers
 $l^\pm = l \pm \half$ and $m^\pm = m \pm \half$. Having in mind that
according to formula (\ref{C-solved}) solutions for positive and negative
$\lambda$ are different we may write (here again $x = \cos \theta$):
\begin{eqnarray}\label{upsP}
  \Upsilon^\pm_{lm} (x,\, \phi) & = &
  \pm i^{l^+} \, (-1)^{\half(m^- +|m^-|)}
    \frac{\sqrt{(l + m)!(l - m)!}}{2^{|m| +\half}\, \Gamma(l^+)}
     \nonumber \\
    & \times &
    \frac{e^{i m \phi}}{\sqrt{2\pi}}\,
    \left(
    \begin{array}{r}
    \sqrt{\mp i \, \rho^{(|m^-|,\, |m^+|)}(x)}\,
    P^{(|m^-|,\, |m^+|)}_{l-|m|} (x)
      \\  \sgn m \,
     \sqrt{\pm i \, \rho^{(|m^+|,\, |m^-|)}(x)}\,
    P^{(|m^+|,\, |m^-|)}_{l-|m|} (x)
    \end{array}\right);
\end{eqnarray}
Here the $\pm$-superscripts stand for the sign of $\lambda$ and
$\rho^{(\alpha,\, \beta)} = (1-x)^\alpha (1+x)^\beta$ is the weight function
for Jacobi polynomials $P^{(\alpha,\, \beta)}(x)$ defined in the Appendix~A.
The overall multiplicative constants in these expressions were chosen in
order to fix the right signs of matrix elements (\ref{UpsLUps}) and ensure
the correct behavior of spherical functions under complex conjugation
(\ref{Ups*}).

The relation between $\Upsilon^+$ and $\Upsilon^-$ is very simple:
\begin{equation}\label{Ups+/Ups-}
  \Upsilon^\pm_{lm} (x,\, \phi) =
    \pm i \sigma_z \, \Upsilon^\mp_{lm} (x,\, \phi).
\end{equation}

The structure of this representation allows to prove directly the
orthogonality of spherical functions:
\begin{equation}\label{onorm}
  \langle \Upsilon^{\varepsilon_1}_{l_1 m_1} |
    \Upsilon^{\varepsilon_2}_{l_2 m_2}\rangle =
    \int_0^{2\pi} d\phi \int_0^\pi
    (\Upsilon^{\varepsilon_1}_{l_1 m_1})^\dagger
    \Upsilon^{\varepsilon_2}_{l_2 m_2}\,
    \sin \theta\, d\theta =
    \delta^{\varepsilon_1 \varepsilon_2}\,
    \delta_{l_1 l_2}\, \delta_{m_1 m_2}.
\end{equation}
Integration over $d\phi$ generates the Kronecker $\delta_{m_1
m_2}$-symbol. After that the integrand becomes a sum of two
expressions of the type
 $\rho^{(|m^\pm|,\,|m^\mp|)}
     P^{(|m^\pm|,\, |m^\mp|)}_{l_1 -|m|}\,
        P^{(|m^\pm|,\, |m^\mp|)}_{l_2-|m|}$
that adverts us to the orthogonality of Jacobi polynomials (\ref{J-ort}). If
the only distinction between the $\Upsilon$-functions is $\epsilon_1 \neq
\epsilon_2$ one should turn back to eqn.~(\ref{Ups+/Ups-}). Due to the
presence of $\sigma_z$ for $\epsilon_1 = - \epsilon_2$ the equal
contributions of upper and lower spinor components are subtracted from each
other and cancel.

Two more economic representations can be found with the help of identities
(\ref{dl+m/dl-m}) of the Appendix. The first is:
\begin{eqnarray} \label{ups>}
  \Upsilon^\pm_{lm} (x,\, \phi) & = &
    \pm \frac{i^{l^+}\, (-1)^{l^-}}{2^{l^+} \Gamma(l^+)}
    \sqrt{\frac{(l + m)!}{(l - m)!}}
    \nonumber \\
    & \times &
    \frac{e^{i m \phi}}{\sqrt{2\pi}}
    \left(
    \begin{array}{r}
     \sqrt{\mp i} (1-x)^{-\frac{m^-}2} (1+x)^{-\frac{m^+}2}\,
    \frac{d^{l-m}}{dx^{l-m}} (1-x)^{l^-} (1+x)^{l^+}  \\
      \sqrt{\pm i} (1-x)^{-\frac{m^+}2} (1+x)^{-\frac{m^-}2}\,
    \frac{d^{l-m}}{dx^{l-m}} (1-x)^{l^+} (1+x)^{l^-}
    \end{array}\right) .
  \end{eqnarray}
The second one differs from it mainly by the sign of $m$.
\begin{eqnarray} \label{ups<}
  \Upsilon^\pm_{lm} (x,\, \phi) & = &
    \frac{i^{l^+} (-1)^{l-m}}{2^{l^+} \Gamma(l^+)}
    \sqrt{\frac{(l - m)!}{(l + m)!}}
 \nonumber \\
    & \times &
    \frac{e^{i m \phi}}{\sqrt{2\pi}}
    \left(\begin{array}{r}
     \mp \sqrt{\mp i} (1-x)^{\frac{m^-}2} (1+x)^{\frac{m^+}2}
      \frac{d^{l+m}}{dx^{l+m}} (1-x)^{l^+} (1+x)^{l^-}
     \\
     \pm \sqrt{\pm i} (1-x)^{\frac{m^+}2} (1+x)^{\frac{m^-}2}
      \frac{d^{l+m}}{dx^{l+m}} (1-x)^{l^-} (1+x)^{l^+}
    \end{array}\right).
\end{eqnarray}
The important feature of these functions is that they are not zero only for
values of $m$ ranging from $-l$ to $l$. Hence each multiplet contains $2l+1$
terms. Indeed once $l= n+|m|$ is a half-integer then $l^+$ and $l^-$ in
their turn must be integers. As a result the products $(1-x)^{l^\pm}
(1+x)^{l^\mp}$ are polynomials of order $2l$ and their derivatives of orders
higher than $2l$ are zero.

These expressions are perfectly suited for acting on them by $\hat{L}_+$ and
$\hat{L}_-$ operators. First let us note that for any $m$
\begin{equation}\label{L:rho}
    \hat{L}_-\, e^{i m \phi} \left(
\begin{array}{c}
  \left[\rho^{(m^-,\, m^+)}\right]^{-\half} \\
  \left[\rho^{(m^+,\, m^-)}\right]^{-\half}
\end{array}\right) =
    \hat{L}_+\, e^{i m \phi} \left(
\begin{array}{c}
  \left[\rho^{(m^-,\, m^+)}\right]^\half \\
  \left[\rho^{(m^+,\, m^-)}\right]^\half
\end{array}\right) = 0.
\end{equation}
If we take $\Upsilon$ in the form (\ref{ups>}) when acting by $\hat{L}_-$
and in the form (\ref{ups<}) when acting by $\hat{L}_+ $ we shall readily
find that
\begin{mathletters} \label{L:Ups}
\begin{eqnarray}
  \hat{L}_-\, \Upsilon_{l,\, m} =
    \sqrt{(l+m)(l-m+1)}\,  \Upsilon_{l,\, m-1} ;  \label{L:Ups/a} \\
  \hat{L}_+\, \Upsilon_{l,\, m} =
    \sqrt{(l+m+1)(l-m)}\,  \Upsilon_{l,\, m+1}.
   \label{L:Ups/b}
\end{eqnarray}
\end{mathletters}
Thus we retain the standard expressions for the matrix elements of
momentum operators,
\begin{mathletters}\label{UpsLUps}
\begin{eqnarray}
  \langle l,\, m-1\,  |\hat{L}_-|\,  l,\, m \rangle & = &
    \langle l,\, m \, |\hat{L}_+|\,  l,\, m-1 \rangle =
    \sqrt{(l+m)(l-m+1)}
     \label{UpsLUps/a}; \\
    \langle l,\, m \, |\hat{L}_z|\,  l,\, m \rangle &  =  & m.
    \label{UpsLUps/b}
\end{eqnarray}
\end{mathletters}
Deducing these relations directly from formulae (\ref{upsP}) would
require engaging but to (\ref{L:rho}) the Jacobi polynomial
differentiation formula (\ref{J-diff}).

\subsection{Complex conjugation and time-reversal \label{c-conj}}

Finally I would like to comment the choice of coefficients in
definition (\ref{upsP}). We have already mentioned that insofar as
the momentum operators (\ref{L-ops}) are diagonal the upper and
lower spinor components of $\Upsilon_{l,\, m}$ belong to different
representations of the $SU(2)$-group. The eigenvalue equation
(\ref{DiracS2}) and normalization condition (\ref{e-norm}) link
the components to each other and fix their ratio (\ref{C/C}) and
absolute values. In the mean time the overall complex phase
remains free. It may be determined from the behaviour of
eigenfunctions under the complex conjugation. The latter in its
own turn is closely related to properties with respect to
time-reversal symmetry in $2+1$-dimensions. Action of the
time-reversal ($T$) transformation onto spinors is described by
the formulae \cite{Landau/Lifshitz}:
\begin{mathletters}\label{T-conj}
\begin{equation}\label{T-conja}
  T :\psi \rightarrow - i \sigma_y\, \psi^*.
\end{equation}
The $- i \sigma_y$ matrix guarantees that the time reversed
spinors belong to the same fundamental representation of $SU(2)$
group and not to the complex-conjugated one. This definition
agrees with the time-reversal properties of ordinary spherical
functions $Y_{l,\, m}$ (see Appendix~B)
\begin{equation}\label{T-conjb}
  T : Y_{l,\, m}
    \rightarrow Y^*_{l,\, -m} = (-1)^{l-m} Y_{l,\, -m}.
\end{equation}
\end{mathletters}
Definitions (\ref{T-conj}) guarantee the equivalence of
integer-momentum representations of $SU(2)$ written both in terms
of spinors and spherical functions.

The same line was pursued by our definitions of
$\Upsilon$-functions.  The idea was that the $T$-transformation
must convert $\Upsilon$'s into themselves:
 $T : \Upsilon_{l,\, m} \rightarrow (-1)^{l-m} \, \Upsilon_{l,\, -m}$.
This relates components of the complex conjugated spinor
 $\Upsilon^*_{l,\, m}$ to those of $\Upsilon_{l,\, -m}$ as follows:
\begin{equation}\label{Ups*}
  T : \Upsilon_{l,\, m} = -i \sigma_y \Upsilon^*_{l,\, m} =
\left(
\begin{array}{r}
  -\beta^*_{l,\, m} \\
  \alpha^*_{l,\, m}
\end{array}\right) = (-1)^{l-m}\,
\left(
\begin{array}{c}
  \alpha_{l,\,- m} \\
  \beta_{l,\,- m}
\end{array}\right).
\end{equation}
One may show that for the functions introduced in the previous
section this is indeed the case. The simplest way to is to take
$\Upsilon_{l,\, m}$ and $\Upsilon_{l,\, -m}$ in different
representations, say (\ref{ups>}) and (\ref{ups<}). A direct
comparison of the coefficients proves (\ref{Ups*}) (mark that
$(-m)^\pm = - (m^\mp)$).

The time-reversal transformation performed twice changes the sign of the
eigenfunctions:
\begin{equation}\label{Tsqr}
  T^2 : \Upsilon_{l,\, m} =
    (-1)^{2l} \Upsilon_{l,\, m} = - \Upsilon_{l,\, m}.
\end{equation}
This also complies with what should be expected from general considerations
for representations with half-integer momentum.

\section{Relation between $\Upsilon$ and $\Omega$ spherical
spinors\label{Ups-vs-Om}}
\subsection{Cartesian coordinates}\label{Decartes}

Now we shall study the relation of the newly found functions
$\Upsilon$ to the standard spherical spinors. In order to do this
we are going to convert $\Upsilon$-functions to Cartesian
coordinates. This requires passing from the two-dimensional
Riemann sphere $S^2$ to the three-dimensional Euclidean space. In
order to access the new dimension we must first add to the two
polar angles $q^1 = \theta $ and $q^2=\phi $ the radial coordinate
 $q^3 =r$. The relations between the 3-dimensional spherical and
Euclidean coordinates are standard:
\begin{equation} \label{3sph}
 x^1 = x = r\sin \theta \cos \phi ;   \quad
 x^2 = y =r\sin \theta \sin \phi ;   \quad
 x^3 = z =r\cos \theta .
\end{equation}
The old metric tensor and  zweibein on the sphere (\ref{zweibein}) must be
substituted by the new spherical metric and dreibein
\begin{equation}\label{dreibein}
 g_{\alpha \beta } = \diag (r^2,\, r^2 \sin^2\theta,\, 1);
    \qquad
 e_\alpha ^a = \diag (r,\, r\sin \theta,\, 1).
\end{equation}
The place of the third Dirac matrix is taken by $\gamma^3 = \sigma_z$.

\subsection{Spinor transformation rules \label{spinor-transform}}

Now we would like to transform our formulae to the Cartesian
coordinates. Note that the transformation includes a rotation of
the dreibein. Accordingly the spinors will change. Let us remind
the corresponding transformation rules.

Local rotations of the dreibein associated with coordinate changes generate
the spin connection. The spinor and dreibein connections depend on each
other so that $\gamma^a$-matrices with dreibein indices stay constant.

In order to find the explicit form of spinor transformation let us
consider an auxiliary local bilinear
 $A^\alpha(q) = \psi^\dagger(q)\sigma^\alpha \psi(q)$ with
$\psi(q)$ being an arbitrary spinor. Let $q^\alpha$ be the old and
$x^\mu$ the new coordinates (there was no reason yet to claim that
$x^m$ were Euclidean). Being a simple contravariant vector $A$
obeys the following transformation law:
\begin{equation}\label{A(q)->A(x)}
  A^\mu(x) = \frac{\partial x^\mu}{\partial q^\alpha}\, A^\alpha (q)
    \quad \mathrm{and} \quad
  A^a =
    e^a_\mu (x)\, \frac{\partial x^\mu}{\partial q^\alpha}\, e^\alpha_b (q) \,
    \psi^\dagger(q)\sigma^b \psi(q),
\end{equation}
where $x$ and $q$ refer to the same spatial point. Take the note
of the two different dreibeins that appear in this expression: the
first one $e^a_\alpha (q)$ refers to the original frame and
$e^b_\mu (x)$ to the new one. The matrix
 $U^a_b =
    e^a_\mu (x)\, \frac{\partial x^\mu}{\partial q^\alpha}\, e^\alpha_b (q)$
is orthogonal $\tilde U\, U = U\, \tilde U = 1$ (where $\tilde U$
denotes the transposed matrix $\tilde U^a_b = U^b_a$). It
describes the local rotation of objects with dreibein indices that
arise from the coordinate transformation. Let us introduce into
equation (\ref{A(q)->A(x)}) an auxiliary unitary $2 \times 2$
matrix $V (q)$ ($V\,V^\dagger = 1$):
\begin{equation}\label{V/V}
  A^a = U^a_b\, \psi^\dagger\sigma^\alpha \psi =
    e^a_\mu (x)\, \frac{\partial x^\mu}{\partial q^\alpha}\, e^\alpha_b (q) \,
    \psi^\dagger(q)\, V
     V^\dagger\, \sigma^b\, V V^\dagger\, \psi(q).
\end{equation}
We shall show shortly that it is always possible to choose $V$ so that it
will compensate the rotation $U$ of the dreibein putting the Pauli matrices
back into the conventional form:
\begin{equation}\label{V-g-V}
  e^a_\mu (x)\, \frac{\partial x^\mu}{\partial q^\alpha}\, e^\alpha_b (q) \,
    V^\dagger (q) \, \sigma^b\, V (q) = \sigma^a.
\end{equation}
Provided that this condition holds the $\sigma$-matrices will stay the same
in the new coordinate frame. The associated transformation rule for spinors
must be:
\begin{equation}\label{psi->V-psi}
  \psi (x) = V^\dagger \, \psi (q),
    \qquad
  \psi^\dagger (x) = \psi^\dagger (q) \, V.
\end{equation}

Let us calculate the $V$-matrix. Similarly to any other rotation the matrix
$U^a_b$ may be written in terms of the adjoint $O(3)$ rotation generators
$\hat{L}$ as:
\begin{equation}\label{Labc}
  \left[\hat{L}^a\right]^b_c = -i\, \epsilon^{abc} .
\end{equation}
Let unit vector $n_c (q)$ point along the axis and $\theta (q)$ be
the angle of local rotation. Then
\begin{equation}\label{U=exp}
  U^a_b (q) = \left[\exp i \hat{L}^c \, n_c (q) \theta(q) \right]^a_b =
    n^a n_b + (\delta^a_b - n^a n_b) \cos \theta +
    i\, n_c \left[\hat{L}^c\right]^a_b\, \sin\theta.
\end{equation}
From the commutation relations of the Pauli matrices it is straightforward
to see that
\begin{equation}\label{V-expL-V}
   \left[\exp i \hat{L}^c \, n_c \theta\right]^a_b \,
    \left(\exp \frac i2 \sigma^c n_c \theta \mathbin{\sigma^b}
    \exp -\frac i2 \sigma^c\, n_c \theta \right) = \sigma^a.
\end{equation}
Therefore the matrices $V$ and $V^\dagger$ compensating the rotation
(\ref{U=exp}) are
\begin{equation}\label{V=exp}
  V(q) = \exp - \frac i2 \sigma^c n_c (q) \theta(q)
    \qquad \mathrm{and} \qquad
  V^\dagger (q) = \exp \frac i2 \sigma^c n_c (q) \theta(q).
\end{equation}
Thus the transformation of spinors is unambiguously determined by the
rotation angle $\theta$ and direction of the axis $n$ which may be found
either from geometrical considerations or algebraically by solving
equation (\ref{V-g-V}).

\subsection{Cartesian realization of $\Upsilon$-functions}\label{Ups/Om}

Now we can apply the theory and find out what do $\Upsilon$-spinors look
like in the ordinary Cartesian coordinates. First we shall derive the
general formulae transforming spinors from the spherical to the orthogonal
frame and then apply them to the spinors in question. Taking for input the
relation between spherical and orthogonal coordinates (\ref{3sph}), the
explicit spherical dreibein (\ref{dreibein}) and the trivial Cartesian
dreibein $e^a_\mu = \delta^a_\mu$ we obtain the rotation matrix
 ($e^a_\mu = \delta^a_\mu$).
\begin{equation}\label{sph-U-Cart}
  U^a_b =
  e^a_\mu (x) \, \frac{\partial x^\mu}{\partial q^\alpha}\,
    e^\alpha_b (q) =
    \left(
\begin{array}{ccc}
  \cos \theta \cos \phi & - \sin \phi & \sin \theta \cos \phi \\
  \cos \theta \sin \phi &   \cos \phi & \sin \theta \sin \phi \\
  - \sin \theta         &           0 & \cos \theta
\end{array} \right).
\end{equation}
This may be decomposed into the product of two successive turns:
\begin{equation}\label{U=eL*eL}
  U^a_b (q) =
    \left[
        \exp - i\, \phi \hat{L}_3 \exp -i \theta \hat{L}_2
    \right]^a_b .
\end{equation}
Geometrically these two rotations align the local axes of
spherical frame with those of Cartesian one. Obviously $\theta$
and $\phi$ are nothing but the corresponding Euler angles. Spinors
must also suffer the same two successive rotations with
$V$-matrices being:
\begin{mathletters}\label{sph-V-Cart}
\begin{eqnarray}
  V =
    \exp \frac{i\sigma_y}2 \theta \,
    \exp \frac{i\sigma_z}2 \phi & = &
    \left(
\begin{array}{rr}
   e^{\frac{i \phi}2} \cos \frac \theta 2
        & e^{- \frac{i \phi}2} \sin \frac \theta 2
    \\
  -e^{\frac{i \phi}2} \sin \frac \theta 2
        & e^{- \frac{i \phi}2} \cos  \frac \theta 2
\end{array}\right) ;
    \label{sph-V-Cart/a} \\
  V^\dagger =
    \exp - \frac{i\sigma_z}2 \phi \,
    \exp - \frac{i\sigma_y}2 \theta  & = &
    \left(
\begin{array}{rr}
  e^{- \frac{i \phi}2} \cos \frac \theta 2
        & -e^{- \frac{i \phi}2} \sin \frac \theta 2
    \\
  e^{\frac{i \phi}2} \sin \frac \theta 2
        & e^{ \frac{i \phi}2} \cos  \frac \theta 2
\end{array}\right).
    \label{sph-V-Cart/b}
\end{eqnarray}
\end{mathletters}

Now we shall perform the $V$-rotation on spinor spherical functions
$\Upsilon$ (the trick is to use representation (\ref{ups>}) when calculating
the upper line and apply (\ref{ups<}) to the lower one):
\begin{equation}\label{V-Ups}
  V^\dagger\, \Upsilon^\pm_{l\, m} =
    \frac 1{\sqrt 2}
    \left(
\begin{array}{c}
    \sqrt{\frac{l+m}{2l}} \, Y_{l^- m^-} \pm
    \sqrt{\frac{l-m+1}{2l+2}} \, Y_{l^+ m^-}
     \\
    \sqrt{\frac{l-m}{2l}} \, Y_{l^- m^+} \mp
    \sqrt{\frac{l+m+1}{2l+2}} \, Y_{l^+ m^+}
\end{array}
    \right) =
    \frac 1{\sqrt 2}
    \left( \Omega_{l,\, l^-,\, m} \mp \Omega_{l,\, l^+,\, m} \right).
\end{equation}

Here $Y_{l\, m}$ are spherical functions (see Appendix~B) and
 $\Omega_{j,\, l,\, m}$ are the ordinary spherical spinors that
realize an alternative representation of spin-$\half$ states on
the sphere \cite{Akhiezer/Berestetsky}. The quantum numbers
characteristic of a spinor $\Omega_{j,\, l,\, m}$ are: the
half-integer total angular momentum $j > 0$, integer orbital
momentum $l = j \pm \half$ and projection of the total momentum
onto the polar axis $m$ (that is again a half-integer). They are
the two-component functions defined by the formulae
\begin{equation}\label{Ojlm}
  \Omega_{l+\half,\, l,\, m} = \left(
\begin{array}{r}
  \sqrt{\frac{j+m}{2j}} \, Y_{l,\, m-\half} \\
  \sqrt{\frac{j-m}{2j}} \, Y_{l,\, m+\half}
\end{array}\right)
    \qquad \mathrm{and} \qquad
  \Omega_{l-\half,\, l,\, m} = \left(
\begin{array}{r}
  - \sqrt{\frac{j-m+1}{2j+2}} \, Y_{l,\, m-\half} \\
  \sqrt{\frac{j+m+1}{2j+2}} \, Y_{l,\, m+\half}
\end{array}\right).
\end{equation}
Similarly to $\Upsilon$ the $\Omega$-spinors make an orthonormal basis in
the space of spinor functions on $S^2$:
\begin{equation}\label{O-ort}
  \int_0^\pi \sin\theta\, d\theta\, \int_0^{2\pi} d\phi\,
    \Omega_{j\, l\, m} (\cos \theta ,\, \phi)\,
        \Omega_{j'\, l'\, m'} (\cos \theta ,\, \phi) =
    \delta_{j\, j'}\, \delta_{l\, l'}\, \delta_{m\, m'}.
\end{equation}

We conclude that in Cartesian coordinates the functions $\Upsilon$ may be
expressed in terms of the ordinary spherical spinors $\Omega$ (see
(\ref{Ojlm})) but do not coincide with those. Representation (\ref{V-Ups})
helps finding the standard quantum numbers of $\Upsilon$-spinors. Inasmuch
as they are superpositions of states with equal values of $j$ and $m$
$\Upsilon$-functions must the same values of total angular momentum
$\hat{J}$ and its $z$-projection $\hat{J}_z$:
\begin{equation}\label{j,m:Ups}
  \hat{\mathbf{J}}^2 \, \Upsilon_{lm} =
    (\hat{\mathbf{L}}_C + \hat{\mathbf{S}}_C)^2 \, \Upsilon_{lm} =
    l (l+1) \, \Upsilon_{lm}
    \qquad \mathrm{and} \qquad
    \hat{J}_z \, \Upsilon_{lm} = m  \, \Upsilon_{lm};
\end{equation}
where $\hat{\mathbf{L}}_C$ and $\hat{\mathbf{S}}_C$ are the Cartesian
operators of vectors of orbital momentum and spin respectively. In the mean
time being composed of states with different orbital momenta $l^+$ and $l^-$
our spinors do not diagonalize $\hat{\mathbf{L}}_C^2$. Still it is
interesting to calculate its average:
\begin{equation}\label{Ups|L|Ups}
  \langle \Upsilon_{lm} | \hat{\mathbf{L}}_C^2 |\Upsilon_{lm} > =
    \half \left[l^+ (l^+ + 1) + l^- (l^- + 1)\right] =
    \left( l+\half \right)^2 = \lambda^2.
\end{equation}
You see that the average value of $\hat{\mathbf{L}}_C^2$ for
spherical spinors $\Upsilon$ is equal to the squared eigenvalue of
spherical Dirac operator. Clearly this is a mere coincidence since
in this representation $-i\hat{\nabla}$ is diagonal whereas
$\hat{\mathbf{L}^2_C}$ is not.

Now it is the time to compare the two types of spinors. Either of
them forms a complete orthogonal functional system and belongs to
the same representation of the $SU(2)$ group. Moreover
$\Upsilon$'s are linear combinations of $\Omega$'s and \emph{vice
versa.\/} The question arises what is the origin of distinction
between them.

The answer is very simple. Both $\Omega$ and $\Upsilon$
diagonalize one more operator but to $\hat{J}$ and $\hat{J}_z$.
These are the square of orbital momentum $\hat{\mathbf{L}^2_C}$
for $\Omega$ and spherical Dirac operator for $\Upsilon$. The
different choice of supplementary operators leads to different
representations.

The ordinary spinors $\Omega$ were obtained from solving in the flat space
the three-di\-men\-sional Laplace equation on spinors
\begin{equation}\label{3dLaplace}
  -\Delta\, \Psi =
    \left[\frac 1{r^2} \hat{\mathbf{L}}_C^2  +
    \frac 1{r^2}\, \partial_r\, r^2\, \partial_r \right]\, \Psi =
    k^2\, \Psi.
\end{equation}
The Laplace operator is diagonal with respect to spin so that the
upper and lower components of $\Psi$ behave as independent
scalars. Separating the radial and angular dependencies one
arrives at solutions with integer values of orbital momentum $l_C
= 0,\, 1,\dots$ Then those are united into doublets with definite
total momentum $j$ and its projection $j_z = m$.

Now the $\Upsilon$-spinors are found by diagonalizing Dirac operator on the
Riemann sphere that is an essentially curved manifold. One may clearly mark
the difference from flat Euclidean space from the curvature term that
manifestly enters the square of Dirac operator (\ref{Dsqr-Lapl}). Besides
the nontrivial spin connection makes even the covariant two-dimensional
Laplace operator (\ref{Dsqr}) on the sphere differ from the angular part of
(\ref{3dLaplace}).

To summarize one may say that as a result $\Omega$-spinors are
better suited for separation of variables in spherically symmetric
problems of non-relativistic quantum mechanics. On the other hand
$\Upsilon$-functions may be useful when one is specially
interested in properties of fermions localized on the sphere.

In conclusion we present for reference the Cartesian realization of Dirac
operator (\ref{Dirac/sph})
 $-i \hat{\nabla}_C = -i\, V^\dagger \hat{\nabla} V$ with $V$-matrices given
by (\ref{sph-U-Cart})
\begin{eqnarray}\label{nabla-C}
  -i \hat{\nabla}_C & = &
    - i\sigma_x \, e^{i \phi \sigma_z}\,
    (\cos \theta\, \partial_\theta - \sin \theta)
    + i\sigma_z\, (\sin \theta\, \partial_\theta + \cos \theta)
    -\frac{i \sigma_y}{\sin \theta}\,
    e^{i \phi \sigma_z}\, \partial_\phi
    \nonumber \\
    & = &
    -i e^{- \frac {i \sigma_z \phi}2}\, \sigma_x \, \partial_\theta \,
    e^{i \sigma_y \theta}\, e^{ \frac {i \sigma_z \phi}2}
    -\frac{i \sigma_y}{\sin \theta}\, e^{i \phi \sigma_z}\, \partial_\phi .
\end{eqnarray}
Calculation of $(-i \hat{\nabla}_C)^2$ is left to the reader as an exercise.

\section*{Conclusion}

We have found the eigenfunctions of Dirac operator on the Riemann
sphere $S^2$. This family of orthonormal two-component spinors may
be used for basis in problems involving fermions in spherically
symmetric fields. The eigensponors are expressed in terms of
Jacobi polynomials. The solutions depend on spherical angles and
are characterized by the following quantum numbers:
\begin{itemize}
   \item eigenvalues of Dirac operator $\lambda$ which are nonzero
   either positive or negative integers ($\lambda=0$ is excluded by
   Lichnerovicz theorem);
   \item half-integer total angular momentum $l = |\lambda| -
   \half$;
   \item projection of angular momentum onto the polar axis
    $m = -l,\dots\, l$ which is obviously a half-integer.
\end{itemize}
The set of solutions corresponding to highest weight $l$ realizes
a $2l+1$ dimensional representation of $SU(2)$ group with matrix
elements being given by the usual formulae for half-integer spin.
The normalization coefficients are chosen so that the properties
of our solutions with respect to complex conjugation and time
reversal are in accord with those of spherical functions.

The eigenspinors of spherical Dirac operator $\Upsilon$ differ
from the ordinary spherical spinors $\Omega$. These two types of
spinors diagonalize the different sets of operators. Therefore
$\Omega$-spinors are listed by another set of quantum numbers. It
includes: half-integer total angular momentum $j$; integer orbital
momentum $l=j\pm\half$; half-integer projection $m$ of angular
momentum onto the polar axis.

The second source of difference is the curvature of $S^2$. The functions
$\Upsilon$ diagonalize the Dirac operator on the Riemann sphere which is a
genuinely curved manifold. Therefore the spin connection is inevitably
present in the equation. On the other hand the conventional spherical
spinors are found by separating the angular variables in the flat
three-dimensional space $R^3$. Obviously the resulting equations contain no
spin connection unless one inserts it by hand.

As a consequence the $\Upsilon$ spherical spinors must be better suited for
problems formulated on the Riemann sphere (see Introduction for examples).
We hope that the explicit and detailed presentation of properties of
$\Upsilon$-spinors will help to use them for practical needs.

In conclusion I would like to thank Maximillian Kreuzer for the
kind hospitality and discussions during the preparation of the
pilot version of the paper. I'm particularly grateful to
Joseph~C.~Varilly for introducing me to the history of the subject
and rendering me excerpts from his own work.

The work was done with partial support from RFBR grants
00--02--17808 and 00--15--96786.

\startappendix
\section{Properties of Jacobi polynomials \label{Jacobi}}

Jacobi polynomials $P^{(\alpha,\, \beta)}_{n} (x)$ ($\alpha,\,
\beta > -1$, $n$ being the order) are classical orthogonal
polynomials that satisfy the equation of hypergeometric type
\cite{Nikiforov/Uvarov, Bateman/Erdelyi}
\begin{equation}\label{J-eq/def}
  \sigma(x)\, y'' + \tau(x)\, y' + \lambda_n\, y =
    \frac 1{\rho(x)} \frac d{dx} [\sigma(x)\, \rho(x)\, y'] + \lambda_n\, y
    =0,
\end{equation}
with the coefficient functions:
\begin{equation}\label{J-c's}
  \rho^{(\alpha,\, \beta)}(x) = (1-x)^\alpha (1+x)^\beta; \qquad
    \sigma(x) = 1-x^2; \qquad
    \tau(x) = \beta - \alpha - (\alpha + \beta + 2) x.
\end{equation}
The constant $\lambda_n$ is equal to
\begin{equation}\label{lambda}
  \lambda_n = n (n + \alpha + \beta + 1).
\end{equation}
The explicit form of  Jacobi polynomials is given by the differential and
integral Rodrigues' formulas:
\begin{mathletters}\label{Rodrigues}
\begin{eqnarray}
  P^{(\alpha,\, \beta)}_{n} (x) & = &
    \frac{(-1)^n}{2^n\, n!} \frac 1{\rho(x)}
    \frac{d^n}{dx^n}\, [\sigma^n (x)\, \rho (x)]; \label{Rodrigues/a}\\
  P^{(\alpha,\, \beta)}_{n} (x) & = &
    \frac{(-1)^n}{2^n\, n!} \frac 1{\rho(x)}
    \frac{n!}{2\pi i}
    \oint \frac{\sigma^n (z)\, \rho (z)}{(z-x)^{n+1}}\, dz .\label{Rodrigues/b}
\end{eqnarray}
\end{mathletters}
The contour of complex integration in the second equation must encircle the
point $x$.

Jacobi polynomials with given $\alpha$ and $\beta$ are orthogonal on the
interval $[-1,\, 1]$ with the weight $\rho^{(\alpha,\, \beta)}(x)$
\begin{equation}\label{J-ort}
  \int^1_{-1}P^{(\alpha,\, \beta)}_{m} (x) P^{(\alpha,\, \beta)}_{n} (x) \,
    \rho^{(\alpha,\, \beta)}(x) \, dx = \delta_{mn}\,
    \frac{2^{\alpha+\beta+1}\,\Gamma(n+\alpha+1)\, \Gamma(n+\beta+1)}%
    {n!\, (2n+\alpha+\beta+1)\, \Gamma(n+\alpha+\beta+1)}.
\end{equation}
Jacobi polynomials form on this interval a complete set and any square
integrable function can be expanded in terms of $P^{(\alpha,\, \beta)}_{n}$.

Adjacent polynomials are related to each other by the differentiation
formula:
\begin{equation}\label{J-diff}
  \frac{d}{dx} P^{(\alpha,\, \beta)}_{n} (x) =
    \half (n+\alpha+\beta+1)\, P^{(\alpha+1,\, \beta+1)}_{n-1} (x).
\end{equation}

In the paper we met Jacobi polynomials with $\alpha = \beta \pm 1$
and used their special properties. The first two identities help
to relate the upper and lower components of spinor spherical
functions (\ref{J-sln's}). Let us prove that for $k\geq 1$
\begin{mathletters}\label{ab-ids}
\begin{eqnarray}
  \left(\frac{k}{1-x} - \frac d{dx}\right)\,
    P^{(k,\, k-1)}_{n} (x) & = &
    \frac{n+k}{1-x}\,P^{(k-1,\, k)}_{n} (x);
    \label{ab-ids/a} \\
  \left(\frac{k}{1-x} + \frac d{dx}\right)\,
    P^{(k-1,\, k)}_{n} (x) & = &
    \frac{n+k}{1+x}\,P^{(k,\, k-1)}_{n} (x).\label{ab-ids/b}
\end{eqnarray}
\end{mathletters}
The proof is based on the integral Rodrigues' formula
(\ref{Rodrigues/b}). Let $C = \frac 1{2\pi
i}\left(-\half\right)^n$ be the numerical factor common to the
both sides of the equations. Then after differentiation we get:
\begin{mathletters} \label{ab-aux}
\begin{eqnarray}
  \left(\frac{k}{1-x} - \frac d{dx}\right)\,
    P^{(k,\, k-1)}_{n} & = &
    \frac{n+k}{1+x} P^{(k,\, k-1)}_{n} -
    \frac{C(n+1)}{(1-x^2)^k}
    \oint \frac{(1-z^2)^{n+k}\, dz}{(z-x)^{n+2}};
\label{ab-aux/a} \\
  \left(\frac{k}{1+x} + \frac d{dx}\right)\,
    P^{(k-1,\, k)}_{n} & = &
    \frac{n+k}{1-x} P^{(k-1,\, k)}_{n} +
    \frac{C(n+1)}{(1-x^2)^k}
    \oint \frac{(1-z^2)^{n+k}\, dz}{(z-x)^{n+2}}.
    \label{ab-aux/b}
\end{eqnarray}
\end{mathletters}
Integrating the rightmost terms by parts opens the way to the desired result
(\ref{ab-ids}).

The second pair of relations was used in deducing representations
(\ref{ups>}, \ref{ups<}). They are valid only for integer $\alpha$
and $\beta$ and for convenience we shall use the notation of
Section~\ref{Upsilon}. (We remind that both $l^\pm = l \pm \half
\geq 0$ and $m^\pm = m \pm \half$ are integers.)
\begin{mathletters}\label{dl+m/dl-m}
\begin{eqnarray}
  \frac{d^{l+m}}{dx^{l+m}} (1-x)^{l^-} (1+x)^{l^+} & = &
    (-1)^{m^-} \frac{(l+m)!}{(l-m)!}
    \frac {\frac{d^{l-m}}{dx^{l-m}} (1-x)^{l^+} (1+x)^{l^-}}
    {(1-x)^{m^+} (1+x)^{m^-}};  \label{dl+m/dl-m/a} \\
  \frac{d^{l+m}}{dx^{l+m}} (1-x)^{l^+} (1+x)^{l^-} & = &
    (-1)^{m^+} \frac{(l+m)!}{(l-m)!}
    \frac {\frac{d^{l-m}}{dx^{l-m}} (1-x)^{l^-} (1+x)^{l^+}}
    {(1-x)^{m^-} (1+x)^{m^+}}  \label{dl+m/dl-m/b}.
\end{eqnarray}
\end{mathletters}
For example let us prove (\ref{dl+m/dl-m/a}) for $m>0$. Note that relation
(\ref{dl+m/dl-m/b}) extends its validity to negative $m$ and \emph{v.~v.\/}
The $l+m$-th derivative of the product is
\begin{equation}\label{dl+m}
  A = \frac{d^{l+m}}{dx^{l+m}} (1-x)^{l^-} (1+x)^{l^+} =
    \sum_{k=0}^{l+m} {l+m \choose k}
    \frac{d^k}{dx^k} (1-x)^{l^-} \frac{d^{l+m-k}}{dx^{l+m-k}} (1+x)^{l^+}.
\end{equation}
Insofar as $l^+,\, l^- $ are integers the derivatives of $(1-x)^{l^-}$  and
$(1+x)^{l^+}$ of orders higher than $l^-$ and $l^+$ vanish. Thus only the
terms with $m^- \leq k \leq l^-$ survive. Introducing a new summation index
$j = k - m^-$ we obtain after expanding the derivatives
\begin{equation}\label{Sum-j}
  A = \sum_{j=0}^{l-m}
    \frac{(-1)^{j+m^-} (l+m)!}{(m^- +j)! (l^+ -j)!}\:
    \frac{l^-! (1-x)^{l-m-j}}{(l-m-j)!} \:
    \frac{l^+ ! (1+x)^j}{j!}.
\end{equation}
Observing that
\begin{equation}\label{d(1-x)-k}
  \frac{(-1)^j l^+ ! }{(l^+ -j)!} (1-x)^{l-m-j} =
    \frac{\frac{d^j}{dx^j}(1-x)^{l^+}}{(1-x)^{m^+}}
    \quad \mathrm{and} \quad
  \frac{l^- !}{(m^- +j)!} (1+x)^j =
    \frac{\frac{d^{l-m-j}}{dx^{l-m-j}} (1+x)^{l^-}}{(1+x)^{m^-}},
\end{equation}
we immediately find
\begin{equation}\label{dl-m}
  A = \frac{(l+m)!}{(l-m)!}
    \frac{(-1)^{m^-}}{(1-x)^{m^+} (1+x)^{m^-}}
     \sum_{j=0}^{l-m} {l-m \choose j}
    \frac{d^j}{dx^j}(1-x)^{l^+} \frac{d^{l-m-j}}{dx^{l-m-j}} (1+x)^{l^-}.
\end{equation}
This is nothing but the RHS of equation (\ref{dl+m/dl-m/a}). The
proof of (\ref{dl+m/dl-m/b}) may be carried out by simply
reversing the sign $x \rightarrow -x$.

\section{Spherical functions \label{Sph-func}}
\setcounter{equation}{0}

Here we cite for reference the basics of spherical functions and
Legendre polynomials.

The associated Legendre polynomials $P^m_l (x_)$ with $l$, $m$
being nonnegative integers are solutions to the second order
ordinary differential equation \cite{Nikiforov/Uvarov,
Bateman/Erdelyi}
\begin{equation}\label{Plm-eqn}
  \frac d{dx}(1-x^2) \frac d{dx}\, y -\frac{m^2}{1-x^2}\, y + l(l+1)\, y = 0.
\end{equation}
The two convenient representations of functions $P^m_l (x_)$ are:
\begin{mathletters}\label{Plm=dl-m/dl+m}
\begin{eqnarray}
  P^m_l (x) & = & \frac{(-1)^l}{2^l\, l!}\, (1-x^2)^{\frac m2}
    \frac{d^{l+m}}{dx^{l+m}}(1-x^2)^l
    \label{Plm=dl-m/dl+m/a} \\
  P^m_l (x) & = &
    \frac{(-1)^{l-m}}{2^l\, l!}\, \frac{(l+m)!}{(l-m)!}\, (1-x^2)^{-\frac m2}
    \frac{d^{l-m}}{dx^{l-m}}(1-x^2)^l .
  \label{Plm=dl-m/dl+m/b}
\end{eqnarray}
\end{mathletters}
They may be also expressed in terms of the Jacobi and Legendre polynomials,
\begin{equation}\label{Plm-Pl-Jacobi}
  P^m_l (x) =
    \frac{(l+m)!}{2^m\, l!}\, (1-x^2)^{\frac m2}\,  P^{(m,\, m)}_{l-m}(x),
    \qquad \mathrm{and} \qquad
  P^m_l (x) = (1-x^2)^{\frac m2}\,\frac{d^m}{dx^m}  P_l (x).
\end{equation}
From the orthogonality of Jacobi polynomials (\ref{J-ort}) it follows that
associated Legendre polynomials with the same value of $m$ are orthogonal on
the interval $[-1, 1]$ with the unit weight:
\begin{equation}\label{Plm-ort}
  \int_{-1}^1 dx\, P^m_k (x)\, P^m_l (x) =
   \delta_{kl}\, \frac{(l+m)!}{(l-m)!}\, \frac 2{2l+1}.
\end{equation}

Spherical functions $Y_{l\,m} (x,\, \phi)$ are listed by two integers,
namely, by the value of the angular momentum $l \geq 0$ and its projection
onto the polar axis $-l \leq m \leq l$. The functions $Y_{l\,m} (x,\, \phi)$
are normalized eigenfunctions of the scalar (spin-zero) Laplace operator on
the sphere:
\begin{equation}\label{LapY=l(l+1)Y}
  \left(
  -\frac 1{\sin\theta}\, \frac \partial{\partial\theta}\,
    \sin\theta\, \frac \partial{\partial\theta} -
    \frac 1{\sin^2 \theta}\, \frac{\partial^2}{\partial\phi^2}
    \right)\, Y_{l\,m} (\cos\theta,\, \phi) =
    l (l+1)\, Y_{l\,m} (\cos\theta,\, \phi).
\end{equation}
They may be expressed in terms of the associated Legendre polynomials:
\begin{equation}\label{Ylm-Plm}
  Y_{l\,m} (x,\, \phi) = (-1)^{\frac{m+|m|}2}\, i^l\,
    \frac{e^{im\phi}}{2 \pi}\, \sqrt{\frac{2l+1}2\, \frac{(l-|m|)!}{(l+|m|)!}}
    P_l^{|m|} (x).
\end{equation}
With the help of identities (\ref{Plm=dl-m/dl+m}) the spherical functions
may be written as
\begin{mathletters} \label{Ylm=dl-m/dl+m}
\begin{eqnarray}
  Y_{l\, m} (x,\, \phi) & = &
    \frac{(-1)^l\, i^l}{2^l\, l!}\,
    \frac{e^{im\phi}}{2 \pi}\,
    \sqrt{\frac{2l+1}2\, \frac{(l+m)!}{(l-m)!}}\,
    (1-x^2)^{-\frac m2} \frac{d^{l-m}}{dx^{l-m}}(1-x^2)^l .
    \label{Ylm=dl-m/dl+m/a}
 \\
  Y_{l\,m} (x,\, \phi) & = &
    \frac{(-1)^{l-m}\, i^l}{2^l\, l!}\,
    \frac{e^{im\phi}}{2 \pi}\,
    \sqrt{\frac{2l+1}2\, \frac{(l-m)!}{(l+m)!}}\,
    (1-x^2)^{\frac m2} \frac{d^{l+m}}{dx^{l+m}}(1-x^2)^l .
    \label{Ylm=dl-m/dl+m/b}
\end{eqnarray}
\end{mathletters}
The functions $Y_{lm}(\cos \theta ,\, \phi)$ form on the sphere the
orthonormal functional basis:
\begin{equation}\label{Ylm-ort}
  \int_0^\pi \sin\theta\, d\theta\, \int_0^{2\pi} d\phi\,
    Y_{l\, m} (\cos \theta ,\, \phi)\, Y_{l'\, m'} (\cos \theta ,\, \phi) =
    \delta_{l\, l'}\, \delta_{m\, m'}.
\end{equation}

Inasmuch as the angular part of 3-dimensional Laplace operator
(\ref{LapY=l(l+1)Y}) coincides with the square of angular momentum
$\hat{\mathbf{L}}_C^2$ in coordinate representation spherical functions
$Y_{l\,m}$ are the basis vectors of $2l+1$ dimensional representation of
rotation group $O(3)$.


\end{document}